\documentstyle [12pt] {article}
\title{A SIMPLE DETERMINATION OF THE THERMODYNAMICAL CHARACTERISTICS OF A VERY THIN BLACK RING}

\author{Vladan Pankovi\'c$^{\ast,\sharp}$,
Simo Ciganovi\'c$^\sharp$\\
$^\ast$Department of Physics, Faculty of Sciences, 21000 Novi
Sad,\\ Trg Dositeja Obradovi\'ca 4. , Serbia, vdpan@neobee.net \\
$^\sharp$Gimnazija, 22320 Indjija, Trg Slobode 2a, Serbia \\}

\date {}
\begin{document}
\maketitle

 PACS number :   04.70.Dy

\begin {abstract}
In this work we suggest a very simple, approximate formalism for
description of some basic (especially thermodynamical)
characteristics of a rotating, very thin black ring. (In fact, our
formalism is not theoretically dubious, since, at it is not hard
to see, it can correspond to an extreme simplification of a more
accurate, Copeland-Lahiri string formalism for the black hole
description.) Even if suggested formalism is, generally speaking,
phenomenological and rough, obtained final results, unexpectedly,
are non-trivial. Concretely, given formalism reproduces exactly
Bekenstein-Hawking entropy, Bekenstein quantization of the entropy
or horizon area and Hawking temperature of a rotating, very thin
black ring obtained earlier using more accurate analysis by Reall,
Emparan, Elvang, Virmani etc. (Conceptually it is similar to
situation in Bohr's atomic model where energy levels are
determined practically exactly even if electron motion is
described roughly.) Our formalism, according to suggestions in our
previous works, is physically based on the assumption that
circumference of the horizon tube holds the natural (integer)
number of corresponding reduced Compton's wave length. (It is
conceptually similar to Bohr's quantization postulate in Bohr's
atomic model interpreted by de Broglie relation.) Also, we use,
mathematically, practically only simple algebraic equations (by
determination of Hawking temperature we use additionally only
simple differentiation of Smarr relation).
\end {abstract}
\vspace {0.5cm}

A rotating black ring in five (and more) dimensions [1], [2]
represents a very interesting form of the black hole whose
studying yields very important conclusions (impossibility of the
simple, five-dimensional generalization of the uniqueness theorem,
possibility of the existence of the naked singularity, etc.). But
detailed analysis of its basic physical characteristics is not
simple. In this work, according to suggestions in our previous
works [3]-[5], we shall suggest a very simple, approximate
formalism for description of some basic (especially
thermodynamical) characteristics of given rotating, very thin
black ring. (In fact, our formalism is not theoretically dubious,
since, at it is not hard to see, it can correspond to an extreme
simplification of a more accurate, Copeland-Lahiri string
formalism for the black hole description [6].) Even if suggested
formalism is, generally speaking, phenomenological and rough,
seemingly "naive", obtained final results, unexpectedly, are
non-trivial. Concretely, given formalism reproduces exactly, for
rotating, very thin black ring, Bekenstein-Hawking entropy,
Bekenstein quantization of the entropy or horizon area and Hawking
temperature, any of which has been obtained earlier using more
accurate analysis (Reall, Emparan, Elvang, Virmani [1], [2]).
(Conceptually it is similar to situation in Bohr's atomic model
where energy levels are determined practically exactly even if
electron motion is described very roughly.) Our formalism is
physically based on the assumption that circumference of the
horizon tube holds the natural (integer) number of corresponding
reduced Compton's wave length. (It is conceptually similar to
Bohr's quantization postulate in Bohr's atomic model interpreted
by de Broglie relation.) Also, we shall use, mathematically,
practically only simple algebraic equations (by determination of
Hawking temperature we shall use additionally only simple
differentiation of Smarr relation).

As it is well-known [1], [2], horizon of a rotating, very thin
black ring can be approximated by a torus with great radius (i.e.
distance from the center of the tube to the center of the torus)
\begin {equation}
    R_{1} = R(\frac {\lambda}{\nu})^{\frac {1}{2}}
\end {equation}
and small radius (i.e. radius of the torus tube)
\begin {equation}
   R_{2} = R\nu
\end {equation}
, where $R$, $\lambda$ and $\nu$ represent corresponding variables
so that, for rotating, very thin black ring, it is satisfied
\begin {equation}
    0 < \nu < \lambda \ll 1     .
\end {equation}
It is well-known too [1], [2] that rotating, very thin black ring
holds (in the natural units system $\hbar = G = c = k = 1$) the
following mass - $M$, angular momentum - $J$, angular velocity -
$\Omega$, horizon area - $A$, entropy - $S$, and temperature - $T$
\begin {equation}
    M = \frac {3\pi}{4}R^{2}\lambda = \frac {3\pi}{4}  R_{1}R_{2} (\frac {\lambda}{\nu})^{\frac {1}{2}}
\end {equation}
\begin {equation}
    J = \frac {\pi R^{3}}{2} \lambda (1 - \frac {\nu}{\lambda}) ^{\frac {1}{2}}= \frac {\pi}{2} R^{2}_{1}R_{2}(1 - \frac {\nu}{\lambda}) ^{\frac {1}{2}}
\end {equation}
\begin {equation}
   \Omega = \frac {1}{R}(1 - \frac {\nu}{\lambda}) ^{\frac {1}{2}}  = \frac {1}{R_{1}}(\frac {\lambda}{\nu}- 1) ^{\frac {1}{2}}
\end {equation}
\begin {equation}
   A = 8 \pi^{2}R^{3}\nu^{\frac {3}{2}}\lambda^{\frac {1}{2}}=  8 \pi^{2} R^{2}_{2}R_{1}
\end {equation}
\begin {equation}
    S = \frac {A}{4}= 2\pi^{2} R^{2}_{2}R_{1}
\end {equation}
\begin {equation}
   T = \frac {1}{4\pi R}\nu^{-\frac {1}{2}}\lambda^{-\frac {1}{2}} = \frac {1}{4\pi R_{2}} (\frac {\nu}{\lambda})^{\frac {1}{2}}
\end {equation}

Now, as well as it has been done in our previous works [3]-[5],
{\it suppose (postulate) } the following expression
\begin {equation}
      m_{+n}R_{2} = \frac {n}{2\pi} \hspace{1cm}   {\rm for}  \hspace{1cm}  n=1,
      2,...
\end {equation}
where $ m_{+n}$ for $n = 1, 2, …$ represent masses of some quantum
systems captured at given rotating, very thin black ring horizon.
It implies
\begin {equation}
      2\pi R_{2} = \frac {n}{m_{+n}} = n \lambda_{r+n} \hspace{1cm}   {\rm for}  \hspace{1cm}  n=1,
      2,...
\end {equation}
Here, obviously, $2\pi R_{2}$ represents the circumference of
small circle at given rotating, very thin black ring horizon,
while
\begin {equation}
      \lambda_{r+n}=\frac {1}{m_{+n}}
\end {equation}
represents the $n$-th reduced Compton wavelength of a quantum
system with mass $ m_{+n}$ captured at very thin black ring
horizon for $n = 1, 2,...$ .

Expression (11) simply means that {\it circumference of given
small  circle (tube) at given rotating, very thin black ring
horizon holds exactly n corresponding} $n$-{\it th reduced Compton
wave lengths of mentioned quantum systems for} $n = 1, 2,...$ .
(Obviously, it is conceptually similar to well-known Bohr's
quantization postulate interpreted by de Broglie relation
(according to which circumference of $n$-th electron circular
orbit contains exactly $n$ corresponding $n$-th de Broglie wave
lengths, for $n = 1, 2, …$).

According to (11) it follows
\begin {equation}
       m_{+n}= \frac {n}{2\pi R_{2}} = n m_{+1} \hspace{1cm}   {\rm for}  \hspace{1cm}  n=1,
       2,...
\end {equation}
where
\begin {equation}
       m_{+1} = \frac {1}{2\pi R_{2}}
\end {equation}

Now, analogously to procedure in [3]-[5], {\it suppose (postulate)
that black hole entropy is proportional to quotient of given
rotating, very thin black ring mass} $M$ {\it and minimal mass of
mentioned quantum systems} $ m_{+1}$. More precisely, suppose
(postulate)
\begin {equation}
      S = \gamma \frac{M}{ m_{+1}} = \gamma (\frac {3 \pi^{2}}{2}) R_{1}R^{2}_{2}(\frac {\lambda}{\nu})^{\frac {1}{2}}
\end {equation}
where $\gamma$ represents some normalization factor, i.e.
parameter that can be determined by comparison of (15) and (8).
(It can be supposed that within given approximation normalization
factor cannot be determined so that it must be obtained by
comparison with exact theory.) It yields
\begin {equation}
      \gamma = \frac {4}{3}(\frac {\nu}{\lambda})^{\frac {1}{2}}
\end {equation}
Also, we shall define normalized mass of given rotating, very thin
black ring
\begin {equation}
      \tilde {M} = \gamma M = \pi R_{1}R_{2}                    .
\end {equation}

Further, we shall differentiate (7), (8) supposing approximately
that $R_{2}$ represents unique variable with relevant
differential. (More precisely, according to (1), (2), for a
rotating, very thin black ring small radius $ R_{2}$ is
significantly smaller than great radius $ R_{1}$. But then
curvature of the small circle (proportional, roughly speaking, to
$ \frac {1}{R_{2}}$ ) becomes significantly larger than curvature
of the great circle (proportional, roughly speaking, to $ \frac
{1}{R_{1}}$). Since, roughly speaking, physical effects are
proportional to curvature, supposition that $ R_{2}$  represents
effectively unique relevant variable for a rotating, very thin
black ring, is relatively consistent. On the other hand, it is
supposed that $\nu$ and $\lambda$ without great and small radius
definition can be considered as parameters without relevant
differentials.) It yields
\begin {equation}
      dA = 16 \pi^{2}R_{1} R_{2} d R_{2}
\end {equation}
\begin {equation}
      dS = 4\pi^{2}R_{1} R_{2} d R_{2}               .
\end {equation}
Since, according to (17),
\begin {equation}
      d\tilde {M} = \pi R_{1}dR_{2}
\end {equation}
that yields
\begin {equation}
      dR_{2} = \frac { d\tilde {M}}{\pi R_{1}}
\end {equation}
, (18), (19) turn out in
\begin {equation}
      dA = 16\pi R_{2} d\tilde {M}
\end {equation}
\begin {equation}
      dS = 4 p \pi R_{2} d\tilde {M}               .
\end {equation}
Given expressions can be approximated by the following finite
difference forms
\begin {equation}
      \Delta A = 16\pi R_{2} \Delta \tilde {M} \hspace{1cm}   {\rm for}  \hspace{1cm}  \Delta \tilde {M}\ll  \tilde {M}
\end {equation}
\begin {equation}
      \Delta S = 4 \pi R_{2} \Delta \tilde {M} \hspace{1cm}   {\rm for}  \hspace{1cm}  \Delta \tilde {M}\ll  \tilde {M}.
\end {equation}
Further, we shall assume
\begin {equation}
   \Delta \tilde {M}= n m_{+1} \hspace{1cm}   {\rm for}  \hspace{1cm}  n=1,
   2,...
\end {equation}
Introduction of (26) in (24), (25), according to (14), yields
\begin {equation}
    \Delta A = 8n = 2n (2)^{2} \hspace{1cm}   {\rm for}  \hspace{1cm}  n=1,
    2,...
\end {equation}
\begin {equation}
    \Delta S = 2n  \hspace{1cm}   {\rm for}  \hspace{1cm}  n=1,
    2,...
\end {equation}
that exactly represent Bekenstein quantization of the black hole
horizon surface (where $(2)^{2}$ represents the surface of the
quadrate whose side length represents twice Planck length, i.e. 1)
and entropy.

Now, we shall determine temperature of given rotating, very thin
black ring in our approximation in the following way. We shall
start from Smarr relation  [1], [2] satisfied for black rings
(including very thin rotating black ring too)
\begin {equation}
   \frac {2}{3}M = TS +  \Omega J
\end {equation}
or, according to (1), (2), (4), (5), (6), (8),
\begin {equation}
  \frac {2}{3}\frac {3\pi}{4}R_{1}R_{2} (\frac {\lambda}{\nu})^{\frac {1}{2}} = T2\pi^{2}R^{2}_{2}R_{1} +  \frac {1}{R_{1}}(\frac {\lambda}{\nu}- 1) ^{\frac {1}{2}}\frac {\pi}{2}R^{2}_{1}R_{2} (1 - \frac {\nu}{\lambda}) ^{\frac {1}{2}}               .
\end {equation}
It, after differentiation for mentioned supposition, i.e.
approximation rule (that only small radius and temperature as a
function of the small radius represent unique relevant variables)
yields
\begin {equation}
  \frac {\pi}{2}R_{1} (\frac {\nu}{\lambda})^{\frac {1}{2}}dR_{2} =  2\pi^{2}R_{1}d(TR^{2}_{2} ) +  \frac {1}{R_{1}}(\frac {\lambda}{\nu}- 1)^{\frac {1}{2}}\frac {\pi}{2}R^{2}_{1}(1 - \frac {\nu}{\lambda})^{\frac {1}{2}}dR_{2}
\end {equation}
or, after simple transformations
\begin {equation}
  (\frac {\nu}{\lambda})^{\frac {1}{2}}(\frac {1}{2}\frac {\lambda}{\nu} - \frac {1}{2} (\frac {\lambda}{\nu}- 1)) dR_{2} =  2 \pi d(TR^{2}_{2})
\end {equation}
i.e.
\begin {equation}
  (\frac {\nu}{\lambda})^{\frac {1}{2}}\frac {1}{2}dR_{2}=  2 \pi  d(TR^{2}_{2}) .
\end {equation}
Then simple integration of (33) yields
\begin {equation}
  (\frac {\nu}{\lambda})^{\frac {1}{2}}\frac {1}{2}R_{2} =  2 \pi (TR^{2}_{2})
\end {equation}
that implies
\begin {equation}
   T = \frac {1}{4\pi R_{2}} (\frac {\nu}{\lambda})^{\frac {1}{2}}
\end {equation}
identical to Hawking temperature of rotating, very thin black ring
(9).

In this way we have reproduced, i.e. determined simply,
approximately three most important thermodynamical characteristics
of rotating, very thin black ring: Bekenstein-Hawking entropy
(15), Bekenstein quantization of the horizon area (27) or entropy
(28), and, Hawking temperature (35). It can be observed, roughly
speaking, that all these characteristics are caused by standing
(reduced Compton) waves at small circles (torus tube circles) at
horizon area only. In other words, within given approximation,
thermodynamical characteristics of given rotating, very thin black
ring are practically independent of the great circle of the torus.

In conclusion it can be shortly repeated and pointed out the
following. In this work we suggested a simple, approximate
formalism for description of some basic (especially
thermodynamical) characteristics of a rotating, very thin black
ring. (In fact, our formalism is not theoretically dubious, since,
at it is not hard to see, it can correspond to an extreme
simplification of a more accurate, Copeland-Lahiri string
formalism for the black hole description.) Even if suggested
formalism is, generally speaking, phenomenological and rough,
obtained final results, unexpectedly, are non-trivial. Concretely,
given formalism reproduces exactly Bekenstein-Hawking entropy,
Bekenstein quantization of the entropy or horizon area and Hawking
temperature of a rotating, very thin black ring obtained earlier
using more accurate analysis (Reall, Emparan, Elvang, Virmani).
Our formalism is physically based on the assumption that
circumference of the horizon tube holds the natural (integer)
number of corresponding reduced Compton's wave length. (It is
conceptually similar to Bohr's quantization postulate in Bohr's
atomic model interpreted by de Broglie relation.) Also, we use,
mathematically, practically only simple algebraic equations (by
determination of Hawking temperature we use additionally only
simple differentiation of Smarr relation).

\vspace{0.5cm}

{\bf References}

\begin{itemize}

\item [[1]] R. Emparan, H. S. Reall, {\it A Rotating Black Ring in Five Dimensions}, hep-th/0110260
\item [[2]] H. Elvang, R. Emparan, A. Virmani, {\it Dynamics and Stability of Black Rings}, hep-th/0608076
\item [[3]] V.Pankovic, M.Predojevic, P.Grujic, {\it A Bohr's Semiclassical Model of the Black Hole Thermodynamics}, Serb. Astron. J., {\bf 176}, (2008), 15; gr-qc/0709.1812
\item [[4]] V. Pankovic, J. Ivanovic, M.Predojevic, A.-M. Radakovic, {\it The Simplest Determination of the Thermodynamical Characteristics of Schwarzschild, Kerr and Reisner-Nordström black hole}, gr-qc/0803.0620
\item [[5]] V. Pankovi\'c, Simo Ciganovi\'c, Rade Glavatovi\'c, {\it The Simplest Determination of the Thermodynamical Characteristics of Kerr-Newman Black Hole}, gr-qc/0804.2327
\item [[6]] E. J. Copeland, A.Lahiri, Class. Quant. Grav. , {\bf 12} (1995) L113 ; gr-qc/9508031

\end{itemize}

\end{document}